\documentclass[aps,prb,onecolumn]{revtex4}
\usepackage{color}\label{key}
\usepackage{soul}
\usepackage{graphicx}
\usepackage{epstopdf}
\usepackage{adjustbox}
\usepackage{array}
\newcolumntype{P}[1]{>{\centering\arraybackslash}p{#1}}
\usepackage{amsmath}

\usepackage [english]{babel}
\begin{document}
\title{High pressure studies on properties of FeGa$_3$: role of on-site coulomb correlation}
\author{Debashis Mondal$^{1,2}$, Velaga Srihari$^3$, C. Kamal$^4$, Himanshu Poswal$^3$, Alka B. Garg$^3$, Arumugam Thamizhavel$^5$, Soma Banik$^1$, Aparna Chakrabarti$^{2,4}$, Tapas Ganguli$^{1,2*}$, and Surinder M. Sharma$^3$}
\address{$^1$Synchrotron Utilization Section, Raja Ramanna Centre for Advanced Technology, Indore, 452013, India.}
\email{tapas@rrcat.gov.in}
\address{$^2$Homi Bhabha National Institute, Anushakti Nagar, Mumbai, 400094, India.}
\address{$^3$High Pressure and Synchrotron Radiation Physics Division, Bhabha Atomic Research Center, Mumbai, 400085, India.}
\address{$^4$Theory and Simulations Lab, HRDS, Raja Ramanna Centre for Advanced Technology, Indore, 452013, India.}
\address{$^5$Tata Institute of Fundamental Research, Homi Bhabha Road, Colaba, Mumbai 400-005, India.}

\begin{abstract}
High pressure X-ray diffraction measurements have been carried out on the intermetallic semiconductor FeGa$_3$ and the equation of state for FeGa$_3$ has been determined. First principles based DFT calculations within the GGA approximation indicate that although the unit cell volume matches well with the experimentally obtained value at ambient pressure, it is significantly underestimated at high pressures and the difference between them increases as pressure increases. GGA + U calculations with increasing values of U$_{Fe(3d)}$ (on-site Coulomb repulsion between the Fe  3d electrons) at high pressures, correct this discrepancy. Further, the GGA+U calculations also show that along with U$_{Fe(3d)}$, the Fe 3d band width also increases with pressure and around a pressure of  4 GPa, a small density of states appear at the Fermi level. High pressure resistance measurements carried out on FeGa$_3$ also clearly show a signature  of an electronic transition. Beyond the pressure of 19.7 GPa, the diffraction peaks reduce in intensity and are not observable beyond $\sim$ 26 GPa, leading to an amorphous state.

\end{abstract}

\maketitle

\section{Introduction}
The study of structural, mechanical, electronic and magnetic properties of intermetallics is an important area of research from both application and fundamental points of view. Although a vast majority of intermetallics are metals, there exist a few semiconductor intermetallic systems also. The values of the electrical band gap in these intermetallic semiconductors strongly depend upon the nature of hybridization between the orbitals of the constituent atoms. A few intermetallic compounds which are reported to have a band gap, include FeSi, FeSb$_2$, RuAl$_2$, FeGa$_3$, RuGa$_3$ and RuIn$_3$. \cite{Jaccarino, Petrovic,Weinert,Amagai,Bogdanov} These are considered to be the promising materials for applications in infrared and thermal devices due to their low band gaps.

In these hybridization induced semiconductors, the hybridization is mainly due to the mixing of d orbitals of transition metal atom with s and p orbitals of the p block elements. Thus, the choice of the elements, their composition and structure of the compounds decide the hybridization strength and eventually the property of the material. The narrow band gap coupled with a large density of states near the valence band edge in these intermetallics plays an important role in increasing their Seebeck coefficients. For example, FeSb$_2$ has a band gap of 0.05 eV and experimental result shows that this material possesses the highest thermoelectric power of -45000 $\mu$V/K. \cite{Bentien} This is quite large as compared to Bi$_2$Te$_3$ (250 $\mu V$/K) based conventional thermoelectric materials. \cite{Goncalves} In case of FeGa$_3$, large negative Seebeck coefficient of 350 $\mu V$/K for single crystal \cite{Hadano} and 563 $\mu V$/K for polycrystalline samples \cite{Amagai} have been measured. Furthermore, it is observed that a small percentage of doping in these intermetallics enhances their thermoelectric power \cite{Kasinathan, Haldolaarachchige, Hadano, Takagiwa} and influences their electronic \cite{Manyala} and /or magnetic properties\cite{Umeo} as well. Along with the existence of a narrow band gap and their associated applications, the transition metal (TM) based intermetallics have also attracted attention because of a more fundamental aspect related to the role of on-site Coulomb correlation in the d band of TM electrons. In some of the systems like FeSi, FeSb$_2$ and  FeGa$_3$, the influence of strong on-site Coulomb correlation on their electronic properties has already been observed. \cite{Fu, Arita2, Herzog}

Among the TM based narrow band semiconductors, FeGa$_3$ has been one of the well studied materials, both experimentally and theoretically. Band gap of this material is found to be in the range of 0.3 - 0.5 eV, which has been obtained by various experimental techniques such as temperature dependent resistivity, temperature dependent magnetic succeptibility, and a combination of photoelectron spectroscopy (PES) and inverse photoelectron spectroscopy (IPES) measurements. \cite{Amagai, Umeo, Tsujii, Arita, Hadano} First-principles based electronic structure calculations with local density approximation (LDA) and generalized gradient approximation (GGA) give similar value for the band gap. \cite{Ulrich2, Imai, Yin, Osorio} The close agreement between the experimental and calculated (using LDA and GGA) values of band gap suggests that the on-site Coulomb correlation may be weak in this system. However, more recently, the results of angle resolved photoelectron spectroscopy (ARPES) measurements reported by Arita \textit{et al.} have shown that there is a large mismatch between the measured and calculated (using LDA) band dispersions at the zone centre. \cite{Arita} They have also performed the electronic structure calculations by using LDA + U approach, where it is found that an on-site Coulomb repulsion of U$ > $3 eV for Fe 3d electrons is necessary to reproduce the band dispersion similar to that obtained from ARPES measurement. It is to be noted that calculation with LDA approximation showed no existence of magnetic moment on FeGa$_3$ rendering it to be non-magnetic. However, in a neutron diffraction measurement \cite{Gamza} and muon spin rotation experiment \cite{Storchak}, existence of finite magnetic moment on Fe atom in FeGa$_3$ has been observed. Theoretical results reported by Yin \textit{et al.} \cite{Yin} have shown that with the incorporation of on-site Coulomb correlation through U within the LDA+U approximation, a magnetic moment on Fe atom is generated in FeGa$_3$.

Though geometric, electronic and magnetic properties of FeGa$_3$ at ambient conditions are reported in literature, there are no experimental studies on the properties of this material under high pressure. To the best of our knowledge, only one computational study on the high pressure properties of FeGa$_3$ is available in literature, which predicts that FeGa$_3$ undergoes a semiconductor to metal transition  at an applied pressure of 25 GPa. \cite{Osorio} The application of external pressure provides a very useful means to modify the nature of bonding and the strength of hybridization between the Fe d and Ga s and p orbitals, resulting in changes in the unit cell volume, electronic and magnetic structures, without introducing any extra chemical element, charge carriers or defects. Comparison of the experimental structural parameters obtained from high pressure X-ray diffraction (XRD) with first-principles calculations also provides a good means to predict the electronic and magnetic properties at high pressure from calculations.

Thus, in order to study and understand the structural properties of FeGa$_3$ at high pressures, we have performed X-ray diffraction measurements on polycrystalline powder sample of FeGa$_3$ under high pressure up to 33 GPa. The equation of state (EOS) for this system has been obtained from the XRD data. We have also carried out electronic structure calculations for this system under high pressure, based on density functional theory (DFT) with generalized gradient approximation (GGA) for exchange-correlation functional. As it is already established from previous studies that on-site Coulomb correlation plays a significant role in determining the electronic properties of FeGa$_3$ at ambient pressure, we have also performed calculations by employing GGA + U approach to account for the effect of on-site Coulomb repulsion in FeGa$_3$. Our analysis shows that at high pressure, the lattice parameters obtained from GGA calculations are significantly underestimated in comparison to the experimental data. Moreover, it has been observed that with increasing pressure, this deviation increases monotonically. We have found that incorporation of on-site Coulomb repulsion on the Fe 3d electrons through U in GGA + U approach, yields a better agreement between the experimental and calculated lattice parameters. In addition, our results suggest that the value of U required to reproduce the experimental lattice parameters, increases with pressure. The present study clearly signifies the importance of U in FeGa$_3$ system at high pressure. However, this does not indicate that the system becomes strongly correlated with an increase in pressure, as our calculations also show that there is also an increase in the bandwidth of Fe 3d electrons due to the  application of pressure. It is also found from the calculations that a small but finite density of states at the Fermi level appears at $\sim $ 4 GPa and beyond, which is also indicated from our electrical resistance measurements.

This paper is organized in the following manner: in the next section, we describe the experimental and computational details employed in the present work. Section III contains the results and discussion, and in Section IV, the conclusion of the work are presented.

\section{Experimental and Computational details}

Polycrystalline FeGa$_3$ powder has been prepared in an induction furnace (frequency 15 kHz) using high purity elements (Fe = 99.98$\%$, Ga = 99.999$\%$ ) in 99.999$\%$ pure Ar gas atmosphere. After preparation, the sample was annealed for 5 days at 600$^{\circ}$C inside a quartz ampule, which was vacuum sealed at $4\times10^{-7}$ mbar pressure. Chemical composition of the sample has been measured by Energy Dispersive Analysis of X-ray (EDAX) measurement. The bulk composition determined from EDAX is Fe$_{0.22}$Ga$_{0.78}$, which is close to our expected composition. Preliminary structural characterization was done by X-ray Diffraction (XRD) measurement at ambient pressure using Cu $K_{\alpha}$ source in a Bruckers Discover D8 system. In order to carry out high pressure XRD measurements, the sample was ground into fine powder. The sample was then wrapped inside a Mo foil and annealed in a vacuum sealed ($4\times10^{-7}$ mbar pressure) quartz ampule at 400$^{\circ}$C for 6 hours, to remove the residual strain developed while grinding.

The binary phase diagram of Fe-Ga reveals that the FeGa$_3$ phase melts incongruently, and hence single crystals cannot be grown directly from its melt. \cite{phaseDia} Hence we adopted the flux growth method to grow the single crystal from a Ga-rich melt composition. The starting materials were of high purity Fe in rod form and Ga taken in the Fe:Ga ratio 1:22, in a high quality recrystallized alumina crucible. The alumina crucible was then subsequently sealed in a quartz ampoule and placed in a resistive heating box type furnace. The temperature of the furnace was raised to 1050$^{\circ}$C and held at this temperature for 24~h to enable proper homogenization. The temperature of the furnace was then cooled down to 600$^{\circ}$C over a period of 3 weeks. The excess flux was centrifuged and the grown crystals were extracted from the crucible. The crystals were of reasonably big size $\sim$ 4~$\times$~5~$\times$~3~mm$^3$.

\begin{figure}
\centering
\includegraphics[width=0.45 \textwidth]{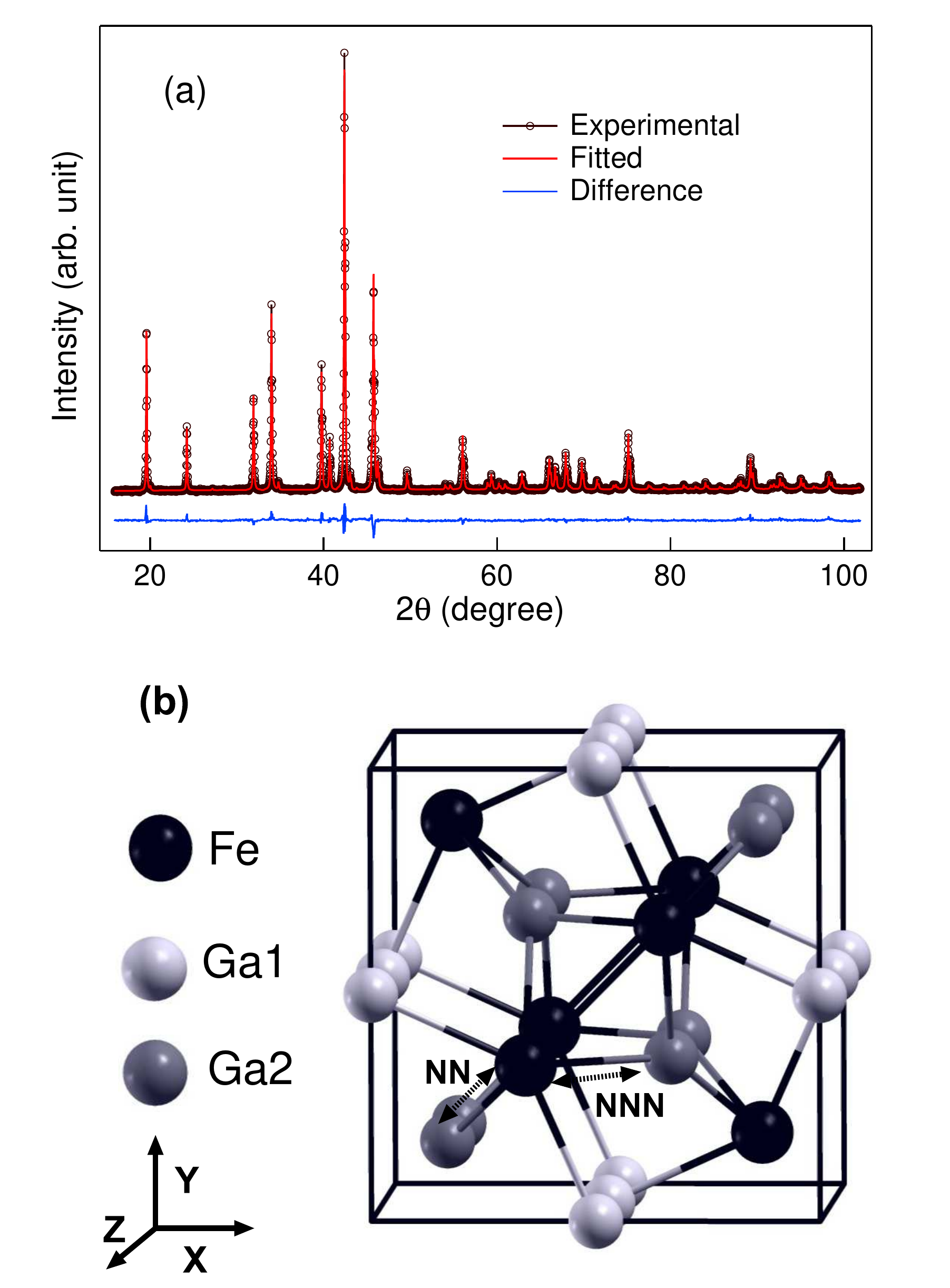}
\caption{(a) X-ray diffraction pattern of FeGa$_3$ powder at ambient pressure using Cu $K_{\alpha}$ source. (b) Ball and stick model of unit cell of FeGa$_3$. The Fe-Fe, Fe-Ga1 and Fe-Ga2 bond lengths are indicated. In the case of Fe-Ga2, the two closely spaced bond lengths are indicated as NN(near neighbor) and NNN(next near neighbor).}
\end{figure}

High pressure XRD measurements have been performed up to  $\sim$ 33 GPa pressure at Extreme conditions XRD (ECXRD) beamline (BL-11), Indus-2, RRCAT \cite{Pandey}, using the angle dispersive mode with a wavelength ($\lambda$) = 0.5692 \AA{}. A MAR345 imaging plate system has been used to collect two dimensional XRD patterns. The calibration of the X-ray photon energy and the distance between the sample and the image plate was carried out by using standard LaB$_6$ and CeO$_2$ samples. FeGa$_3$ and a small amount of Au (used as a pressure marker) was loaded inside a 150 $\mu m$ hole drilled in a pre-indented tungsten gasket (of 40 $\mu m$ thickness) of a diamond anvil cell. A methanol$-$ethanol mixture with a ratio of 4:1 was used as the pressure transmitting medium and the pressure inside the cell was determined by using the known equation of state of gold. To ascertain the consistency of the observation, we have repeated the measurement on FeGa$_3$ twice.

We have also performed high pressure resistance measurement on FeGa$_3$ single crystal by four probe method from ambient to 9 GPa pressure. This measurement on a single crystal sample (2mm$\times$1.5mm$\times$0.1mm) has been carried out in an opposed Bridgman anvil device. A pyrophyllite gasket of thickness 200 $\mu m$ with a central hole of 3 $mm$ in diameter has been used to contain the sample. Bismuth was used for the pressure calibration along with steatite as a pressure transmitting medium. For four probe resistance measurements, stainless steel wires with a diameter of 40 $\mu m$ have been used. A constant current of 1 $mA$ was passed through two outer leads by Keithley source meter and voltage drop across inner two leads was measured using Keithley nanovoltmeter at each value of pressure with two minutes of pressure soaking time.

To study and understand the variation of geometric and electronic structures of FeGa$_3$ under pressure, we have performed density functional theory (DFT) based electronic structure calculations by using Vienna Ab-initio Simulation Package (VASP) within the framework of the projector augmented wave (PAW) method. \cite{Kresse1, Kresse2} For exchange-correlation functional, we have used the generalized gradient approximation (GGA) given by Perdew, Burke and Ernzerhof (PBE). \cite{Perdew} The energy cut-off of plane waves (basis set) was chosen to be 400 eV. For Brillouin zone integration, we have used Monkhorst-Pack scheme with $k$-meshes of $11\times 11\times 10$. The convergence criteria in SCF cycle has been chosen to be $10^{-6}$ eV. The geometric structures have been optimized by minimizing the forces on individual atoms with the criterion that the total force on each atom is below 10$^{-2}$ eV/\AA{}. To incorporate the effects of strong on-site Coulomb repulsion, present in the Fe 3d electrons, we have also carried out similar calculations by employing GGA + U approach using rotationally invariant LSDA+U method \cite{Liechtenstein}.  In this work, we have used U$_{Fe(3d)}$ = 1 to 6 eV for the 3d orbitals of Fe atoms.

\section{Results and discussion}
Figure 1 (a) shows the XRD pattern of FeGa$_3$ obtained using Cu $K_{\alpha}$ source. The results obtained from Rietveld refinement of the XRD data confirm that our sample has tetragonal crystal structure with {\it a} = 6.267 \AA{} and {\it c} = 6.560 \AA{} with $ P4_2/mnm$ space group. The ball and stick model of this system is shown in Fig. 1 (b). The Wyckoff positions of Fe and Ga atoms, obtained from the Rietveld refinement of our data are shown in Table \textrm{I}. These parameters match quite well with the values reported in the literature. \cite{Arita,Ulrich2, Imai, Yin}

\begin{figure}
\centering
\includegraphics[width=0.45 \textwidth]{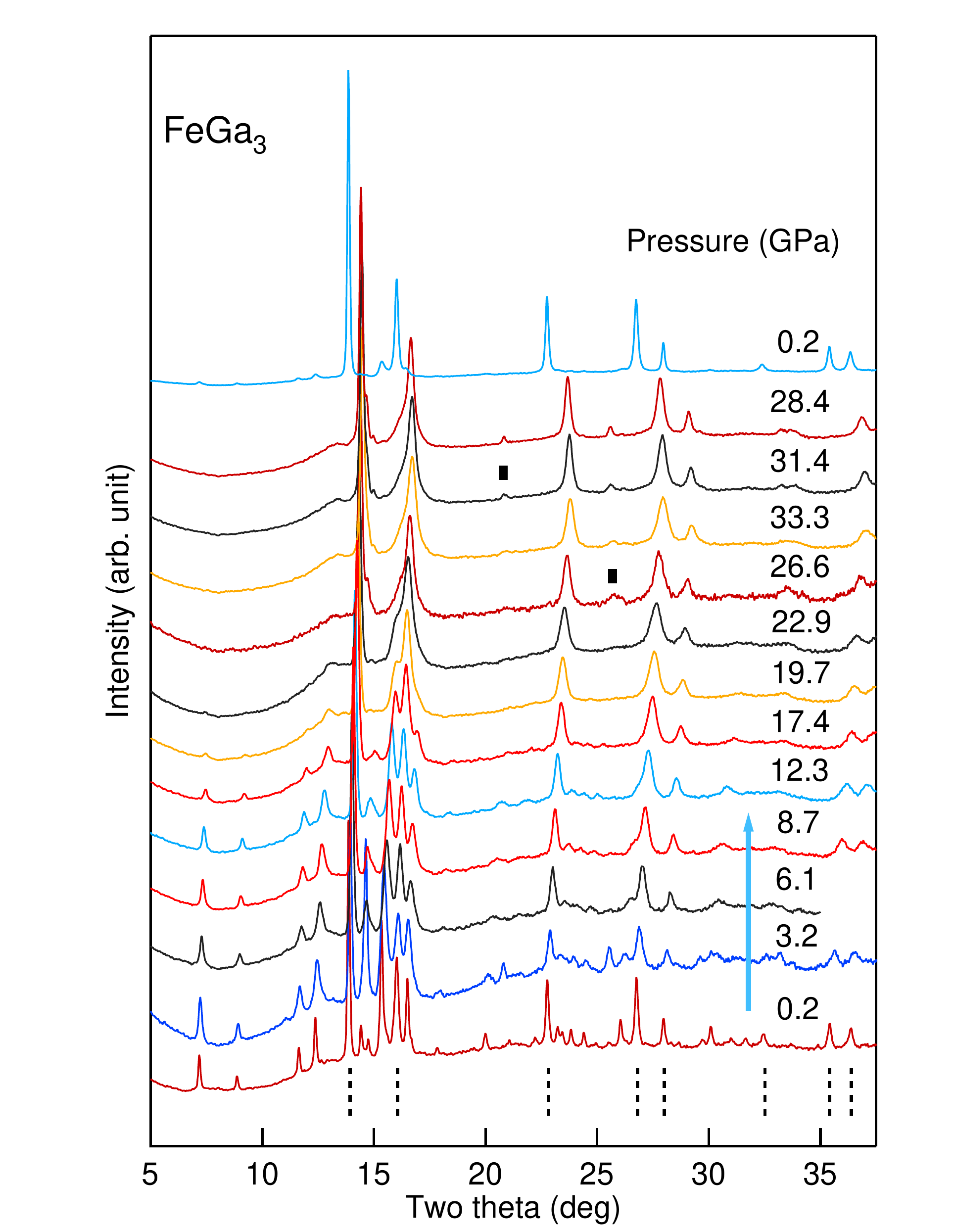}
\caption{Angle-dispersive X-ray diffraction patterns from FeGa$_3$ powder at various applied pressures. The arrow mark shows the direction of the progress of the experiment. Dotted lines and solid black marker indicate the peak positions of Au, originated from pressure marker and the W (tungsten), originated from the gasket respectively.}
\end{figure}

The XRD patterns at various pressures are shown in Fig. 2. In this figure, Au peak positions are marked by dashed lines and the arrow indicates the direction of progress of the experiment. At some pressures, diffraction peaks from tungsten gasket are observed. These are indicated by solid black markers in Fig. 2.   During the high pressure XRD measurements, no new peaks corresponding to new crystalline phase were detected. However, with increase of pressure a clear shift of the diffraction peaks corresponding to the planes in FeGa$_3$ has been observed. Upto a pressure of 19.7 GPa, the peak intensity corresponding to FeGa$_3$ is strong enough to evaluate the  lattice parameters by Le-Bail fitting, but beyond 19.7 GPa, the intensity of the peaks reduces drastically and eventually vanishes at a pressure of 26.6 GPa. This indicates that significant disorder sets in around this pressure, and  the system starts amorphizing at a pressure beyond  $\sim$ 20 GPa. We thus discuss the results in two parts; (a) from ambient pressure to a pressure of  $\sim$ 20 GPa, where the structure is crystalline and Le-Bail fitting could be done; and (b) beyond  $\sim$ 20 GPa, where the system becomes disordered and amorphizes. As the pressure is released, the system again becomes crystalline with tetragonal structure at ambient pressure.

\begin{table}
\vspace{0.5cm}
\caption {Wyckoff positions of the atoms in FeGa$_3$ as determined from the Rietveld refinement of the powder XRD pattern.}
\begin{tabular}{P{1 cm}P{1.4 cm}P{2.0 cm}P{2.0 cm}P{1.6 cm}}
\hline
\hline
Atom	&		Wyckoff position &	x	&	 y	&	z	\\
\hline
\hline
Fe		&	4f&		0.34482(24)&		0.34482(24)&		0\\
Ga1		&	4c&		0&			0.5&		0\\
Ga2		&	8j&		0.15616(12)&		0.15616(12)&		0.26217(17)\\
\hline
\end{tabular}
\end{table}

\subsection{Ambient pressure to  $\sim$ 20 GPa pressure}

We have evaluated the lattice parameters {\it a} and {\it c} at different pressures and consequently the unit cell volume by Le-Bail fitting of the XRD patterns up to 19.7 GPa. The variations of unit cell volume with pressure are shown in Fig. 3. From this figure, we find that there is a monotonic decrease in the unit cell volume with applied pressure.

\begin{figure}
\centering
\includegraphics[width=0.45 \textwidth]{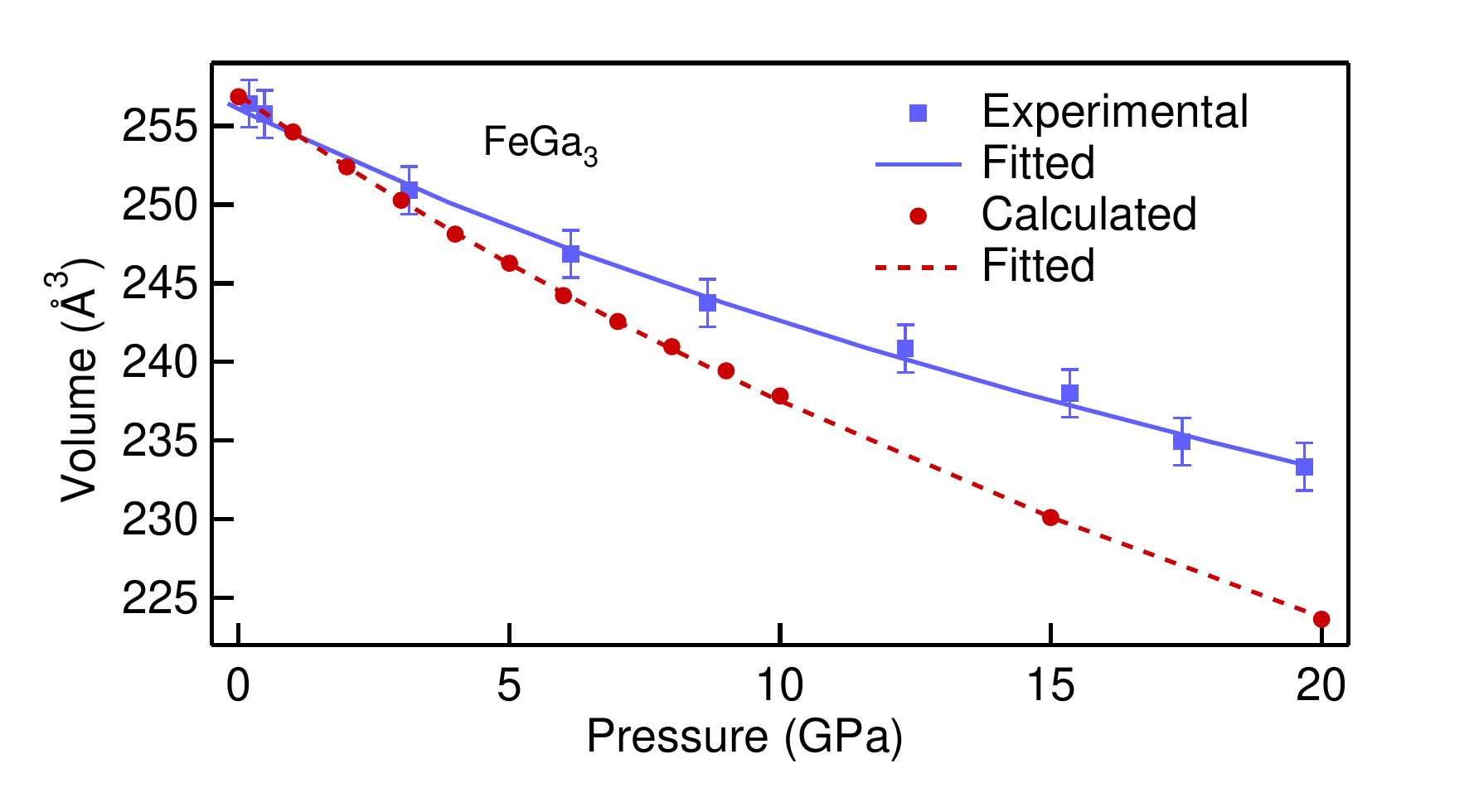}
\caption{Experimental and calculated EOS for FeGa$_3$. Squares and dots represent the experimental and calculated data respectively and lines represent the corresponding fitting with third order Birch-Murnaghan EOS.}
\end{figure}

We have performed DFT based ab-initio calculation within the GGA approximation. The computational details have been discussed in the previous section. The calculated change in the unit cell volume with pressure is also shown in Fig. 3 along with the experimental results. Comparison of experimental and calculated results shows that although the experimental and calculated  values of the cell volume {\it V} are very close ($\Delta V/V $ = 0.15\%) at atmospheric pressure, with an increase in pressure, the calculated values of the unit cell volume are consistently smaller than the experimental values. Further, this difference between the experimental and the calculated value increases monotonically with increase in pressure.

We have fitted the experimental and calculated data with third order Birch-Murnaghan equation of state


\begin{equation}
P(V) = \frac{3B_0}{2}\left[{\left(\frac{V_0}{V}\right)^{7/3}}-{\left(\frac{V_0}{V}\right)^{5/3}}\right]\\\left[1+\frac{3}{4}\left({{B^{\prime}}_0}-4\right)\left[\left(\frac{V_0}{V}\right)^{2/3}-1\right]\right]
\end{equation}

where, $B_o$ is the bulk modulus at ambient pressure of FeGa$_3$, ${{B^{\prime}}_0}$ is first order derivative of $B_o$ with respect to pressure and $V_0$ is the volume of unit cell at ambient pressure. The experimental and calculated values of $B_o$ are 140 GPa and 103 GPa respectively, ${{B^{\prime}}_0}$ are 6.60 and 4.34 respectively, and $V_0$ are 259.0 and 259.4 \AA$^3$ respectively. The difference in the calculated and experimental unit cell volume ($\Delta V/V$) is only 0.15\%, and our calculated bulk modulus at ambient pressure matches very well with earlier calculated report by Osorio-Guillen \textit{et al.} \cite{Osorio}.

To understand and reconcile the mismatch of experimental unit cell volume at high pressure with the calculated values, we have carried out the structural optimization of FeGa$_3$ using GGA+U method with U$_{Fe(3d)}$ = 1, 2, ... 6 eV for Fe 3d orbitals for each value of the pressure (P = 0, 4, 9, 15 and 20 GPa) and compared the unit cell volume with the experimental values. This result has been illustrated in Fig. 4(a). We find that at a pressure of around 4 GPa, the calculated unit cell volume for U$_{Fe(3d)}$ = 3 eV is close to the experimental value and for 9, 15 and 20 GPa the calculated unit cell volume is close to experimental volume for  U$_{Fe(3d)}$ $\sim$ 4, 5, and 6 eV respectively (closest data points are highlighted by black circle in Fig. 4(a)). For clarity we have plotted a 3D graph of difference in unit cell volume between experiment and calculation (along Z axis), at different external pressure (along Y axis) evaluated at various values of U (along X axis). A, B, C and D are the points where this difference in the unit cell volumes is minimum. It shows that as pressure increases, the value of U$_{Fe(3d)}$ for which the difference (between experiment and calculation) in unit cell volume is minimum, also increases. Figure 4(a) also shows that there is no significant difference in the unit cell volume for calculations made with U$_{Fe(3d)}$ = 0 to 3 eV  in the pressure range of our work (ambient to 20 GPa), but it differs significantly beyond 3 eV. Our analysis thus clearly suggests the importance of U$_{Fe(3d)}$ in this system at high pressures. The role of U, within the DFT+U level of calculations in determining the equilibrium lattice parameters has already been reported in several correlated 3d and 4f systems. We give a few examples here. At ambient pressure, in CeO$_2$, it is reported that LDA calculations underestimate the lattice parameters. However, in LDA + U calculations with U = 5\textendash 6 eV, the match between the calculated and experimental lattice parameters is good. \cite{Loschen} In high pressure data on  LaMnO$_3$, it is shown that LDA+U calculations give better bond length estimation as compared to LDA. \cite{Trimarchi} Another report on FeO and MnO also shows that  LDA+U calculations give a better matching with experimentally obtained structural parameters at high pressure as compared to LDA calculations. \cite{Fang} The lattice parameters and structural distortion in NiO with pressure have been studied using GGA+U calculations by Zhang \textit{et al.} \cite{Weibing}. It is shown that, whereas GGA calculations overestimate the structural distortion, GGA+ U calculations give much better results. By comparing the calculated EOS with experimental data for Americium, it was observed that different values of U were necessary in different pressure ranges. \cite{Asok} For TiO$_2$ polymorph, it has been reported that a better description of the crystal and electronic structures requires U $<$ 5 eV, and GGA+U method also gives a correct prediction of phase stability at high pressure. \cite{Arroyo}

\begin{figure}
\centering
\includegraphics[width=0.8 \textwidth]{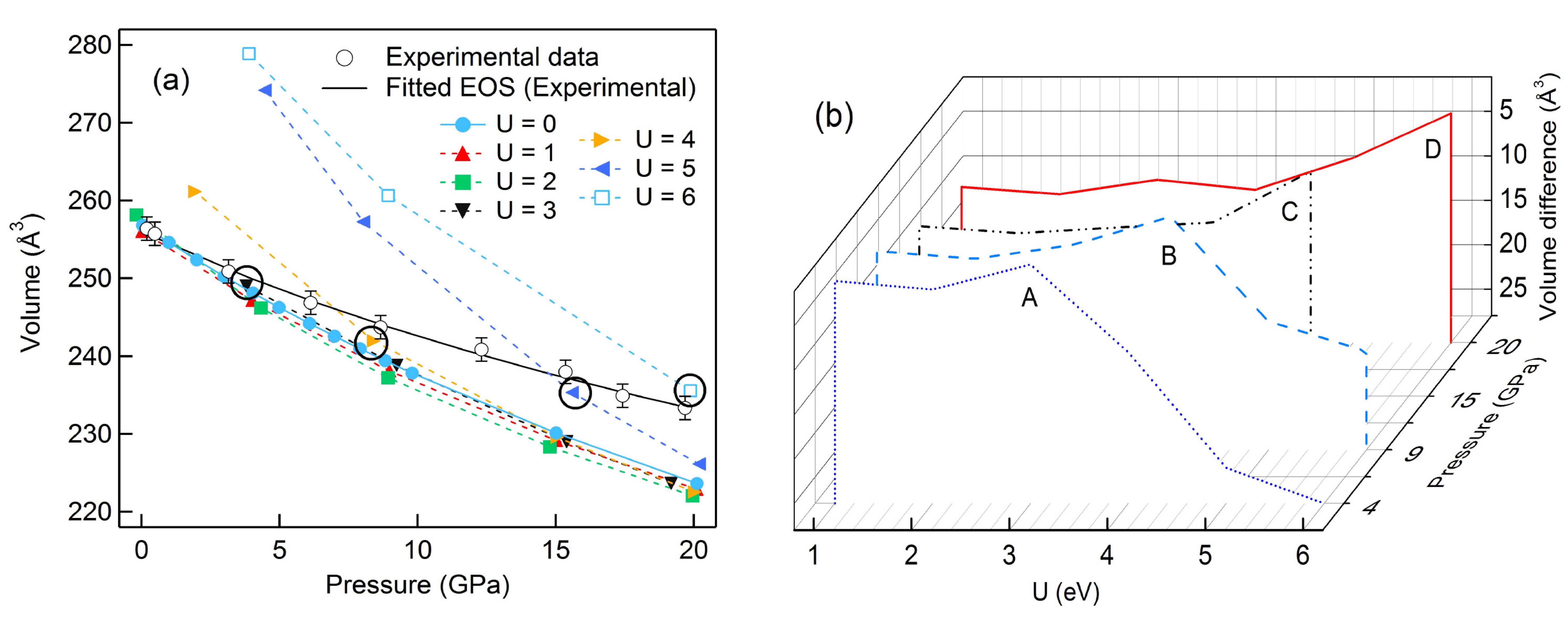}
\caption{(a)The unit cell volume of FeGa$_3$ obtained from the GGA + U calculation for different U$_{Fe(3d)}$ values. Line represents the fitting of experimental data with Birch-Murnaghan's EOS. Black circled data are the closest to the experimentally obtained data at that pressure. (b) a 3D plot of the difference of the volume (along Z) obtained from the experiments and calculations, with respect to pressure (along X) and U$_{Fe(3d)}$ (along Y). For A, U$_{Fe(3d)}$= 3 eV; for B, U$_{Fe(3d)}$= 4 eV; for C, U$_{Fe(3d)}$= 5 eV and for D, U$_{Fe(3d)}$= 6 eV.  In both figures `U' stands for `U$_{Fe(3d)}$'. }
\end{figure}

Further, the role of finite U$_{Fe(3d)}$ at ambient pressure in FeGa$_3$ has been already reported in some recent experiments. Arita \textit{et al.} have carried out ARPES study of single crystal FeGa$_3$ and have noticed that a better agreement of the experimentally obtained dispersion curves with LDA + U calculations for U$_{Fe(3d)}$ = 3 eV is obtained.  \cite{Arita} Results of calculations reported by Yin \textit{et al.} show that within the LDA + U approximation, the band gap of FeGa$_3$ is nearly unchanged for U$_{Fe(3d)}$ = 0 to 2 eV. \cite{Yin, Anita} They have also observed that the d bands are almost flat with very small dispersion in k space, suggesting that the d electrons are localized with high effective mass. Therefore the on-site Coulomb repulsion plays an important role. Yin \textit{et al.} have also reported a small magnetic moment ($\sim$ 0.62 $\mu_B$) on the Fe atom at U$_{Fe(3d)}$ = 2 eV. Recently in a neutron diffraction measurement \cite{Gamza} and muon spin rotation experiment \cite{Storchak}, existence of finite magnetic moment on Fe atom in FeGa$_3$ has been observed. All these reported experimental and theoretical results indicate a significant role of on-site Coulomb repulsion parameter U and consequently a substantial value of U$_{Fe(3d)}$ at ambient pressure.

It is also worth mentioning at this stage that the earlier reports by Hausserman \textit{et al.} \cite{Ulrich2} and Imai \textit{et al.} \cite{Imai}, where the calculations have been carried out with U$_{Fe(3d)}$ = 0, to obtain the band gap and the lattice parameters of FeGa$_3$, a good matching has been observed between experiments and calculations. However recent neutron diffraction measurement \cite{Gamza}, muon spin resonance \cite{Storchak} and our high pressure XRD data cannot be explained without the consideration of significant on-site Coulomb correlation. This apparent contradiction with respect to earlier reports by Hausserman \textit{et al.} \cite{Ulrich2} and Imai \textit{et al.}, \cite{Imai} can be easily resolved by the fact that our GGA + U calculations of the band gap and lattice parameter at ambient pressure do not show any significant variation upto U$_{Fe(3d)}$ $\leq$ 2 eV.

\begin{figure}
\centering
\includegraphics[width=0.6 \textwidth]{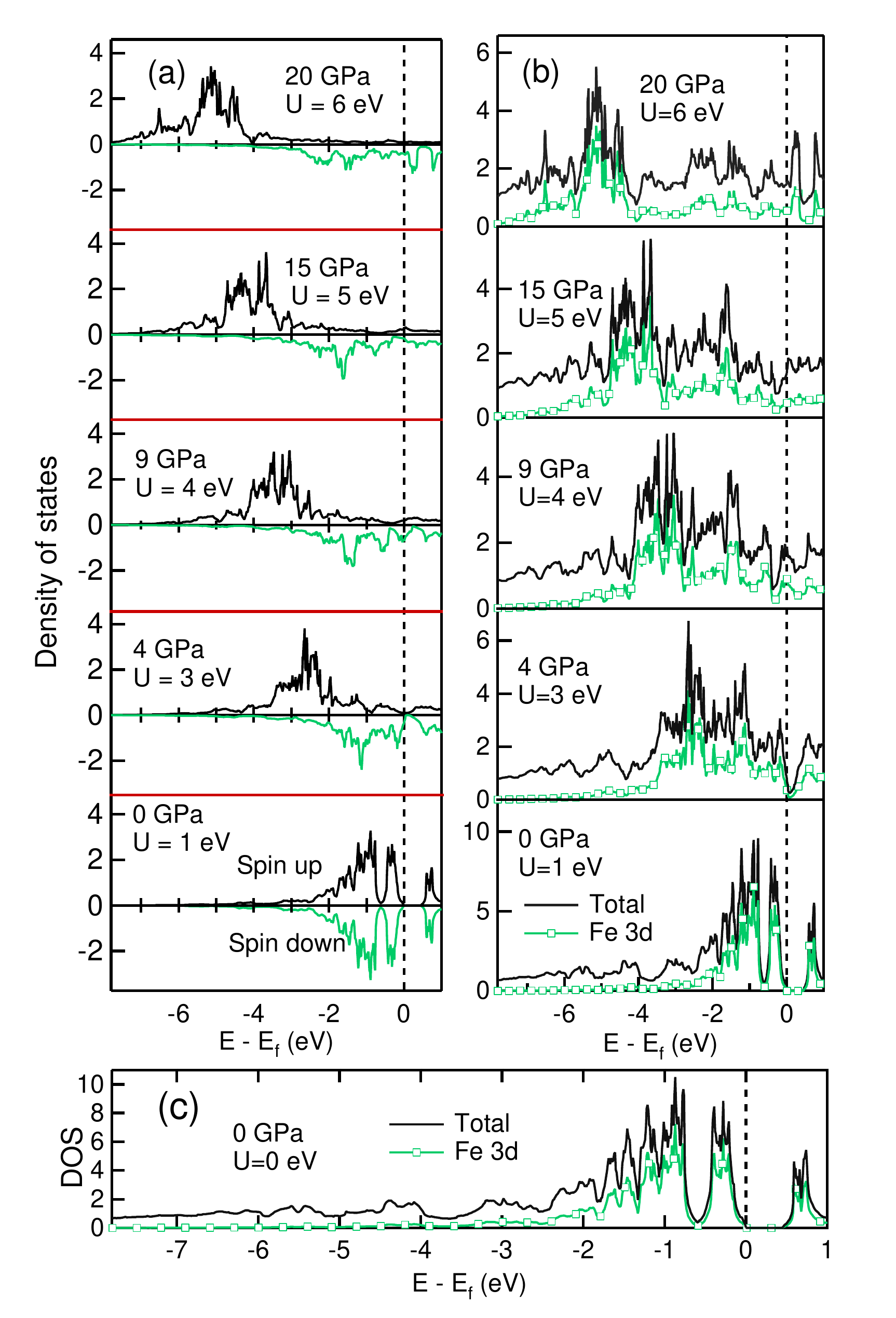}
\caption{Fe 3d partial DOS calculated with GGA and GGA+U approximation. Calculated PDOS are shown for those U$_{Fe(3d)}$ values, for which the calculated and experimental unit cell volume are the closest (a) Spin up and Spin down PDOS of Fe 3d at different pressure (b) Sum of spin up and spin down PDOS of Fe 3d states and total DOS (black colored) of FeGa$_3$ at different U$_{Fe(3d)}$ (c) Fe 3d PDOS and total DOS (black colored) in FeGa$_3$ at ambient pressure with U$_{Fe(3d)}$ = 0 eV. In figure `U' stands for `U$_{Fe(3d)}$'.}
\end{figure}

To find out the evolution of electronic structure in FeGa$_3$ with pressure, we have plotted Fe 3d partial density of states calculated with different U$_{Fe(3d)}$ values (required to reproduce experimental volume at different pressure) resolving into spin up and spin down bands in Fig. 5(a). The total and Fe 3d partial DOS is shown in Fig. 5(b).  For comparison, the total and Fe 3d partial DOS calculated at ambient pressure with U$_{Fe(3d)}$ = 0 eV is also shown in Fig. 5(c). This matches with earlier reported electronic structure calculation quite well. \cite{Arita, Yin, Osorio, Ulrich2, Imai}  We find from Fig 5(a) that, the bandwidth of Fe 3d spin up state increases from  $\sim$ 3 eV at ambient pressure to  $\sim$ 5.4 eV at 20 GPa, and the band width of Fe 3d spin down state remain almost similar ($\sim$ 3 eV) throughout the whole pressure range. Along with the change in the band width, Fe 3d spin up band moves away from the Fermi level and the partial DOS of Fe 3d spin down band decreases near the Fermi level. In this process the total band width of Fe 3d band increases from $\sim$ 3 eV at ambient pressure to $\sim$ 8 eV at 20 GPa.

To determine the role of U$_{Fe(3d)}$ at a given pressure and the role of pressure at a fixed U$_{Fe(3d)}$ independently, we have carried out electronic structure calculations for two cases (a) at different pressure with U$_{Fe(3d)}$ = 0 eV and (b) for different U$_{Fe(3d)}$ values at ambient pressure. The results of these calculations are plotted in Fig. 6 and Fig. 7 respectively. We have plotted total DOS and partial DOS (PDOS) of Fe (3d, 4s, 4p), Ga1 (4s, 4p) and  Ga2 (4s, 4p) atoms along with Fe 3d spin up and spin down band separately. Fig. 6 shows that for U$_{Fe(3d)}$ = 0, all bands broaden uniformly with pressure, maintaining the features in the DOS nearly identical and the bandgap of the system gradually decreases. In case of Fe 3d spin up and spin down bands (shown in Fig. 6(c)), we have not found any spin imbalance upto the maximum pressure of our calculation. In sharp contrast to this, we find in Fig. 7 that the bands modify drastically as U$_{Fe(3d)}$ increases.  With an increase in U$_{Fe(3d)}$, Fe 3d spin up band moves away from the Fermi level and the PDOS of Fe 3d spin down band reduces near the Fermi level which create a spin imbalance (shown in Fig. 7(c)). Due to the decrease in Fe 3d PDOS near the Fermi level (shown in Fig. 7(b)), the hybridization between Fe 3d and Ga s-p states reduces. This reduced hybridization creates relatively pure (metallic like) Ga 4s-4p states near the Fermi level and intensity of these states increases as U increases (shown in Fig. 7(f-i)). We have also found a finite density of states arises at the Fermi level for a U = 3.5 eV.

We find that the value of on-site coulomb repulsion U$_{Fe(3d)}$ (required to reproduce  experimental volume at different applied pressure) varies from 6 eV at 20 GPa  to 3 eV at 4 GPa and  U$_{Fe(3d)}$ is reported to be $\sim$ 2-3 eV at atmospheric pressure in literature as explained above. This corresponds to an increase by a factor between 2 and 3. This observed increase in the value of U$_{Fe(3d)}$ with pressure does not necessarily indicate that there is an increase in the correlation between Fe 3d electrons in FeGa$_3$ with pressure. The other important factor is the variation of the bandwidth ($\Delta$) with pressure that is related to electron de-localization, and $\frac{U}{\Delta}$ is the relevant parameter that determines the importance of correlation. \cite{Tsuchiya} The Fe 3d bandwidth also increases from  $\sim$ 3 eV at ambient pressure to  $\sim$ 8 eV at 20 GPa (shown in Fig. 5(a)), which is an increase of a factor of  $\sim$ 2.5. The significance of U$_{Fe(3d)}$ at high pressure can be understood from the physical argument that with the application of pressure, the atoms come closer to each other, and the orbitals of atoms are  compressed. As spatial extent of the orbitals is reduced, the on site coulomb repulsion on Fe 3d electrons is increased. Simultaneously, as the orbitals come closer, the overlap between Fe 3d and Ga 4p orbitals also increase. This increase in overlap increases the bandwidth, thereby enhancing the itinerancy of the Fe 3d electrons. These two competing effects lead to a value of $\frac{U}{\Delta}$ which remains nearly constant throughout the pressure range studied in this work. We find that in FeGa$_3$, $\frac{1}{\Delta}\frac{d\Delta}{dP}$( $\sim$ 0.083) is close to $\frac{1}{U}\frac{dU}{dP}$ ( $\sim$ 0.1), which indicates that, at high pressure, increase in on-site Coulomb repulsion U$_{Fe(3d)}$ and itinerancy are of the same magnitude and hence equally important in this system.

\begin{figure}
\centering
\includegraphics[width=0.9 \textwidth]{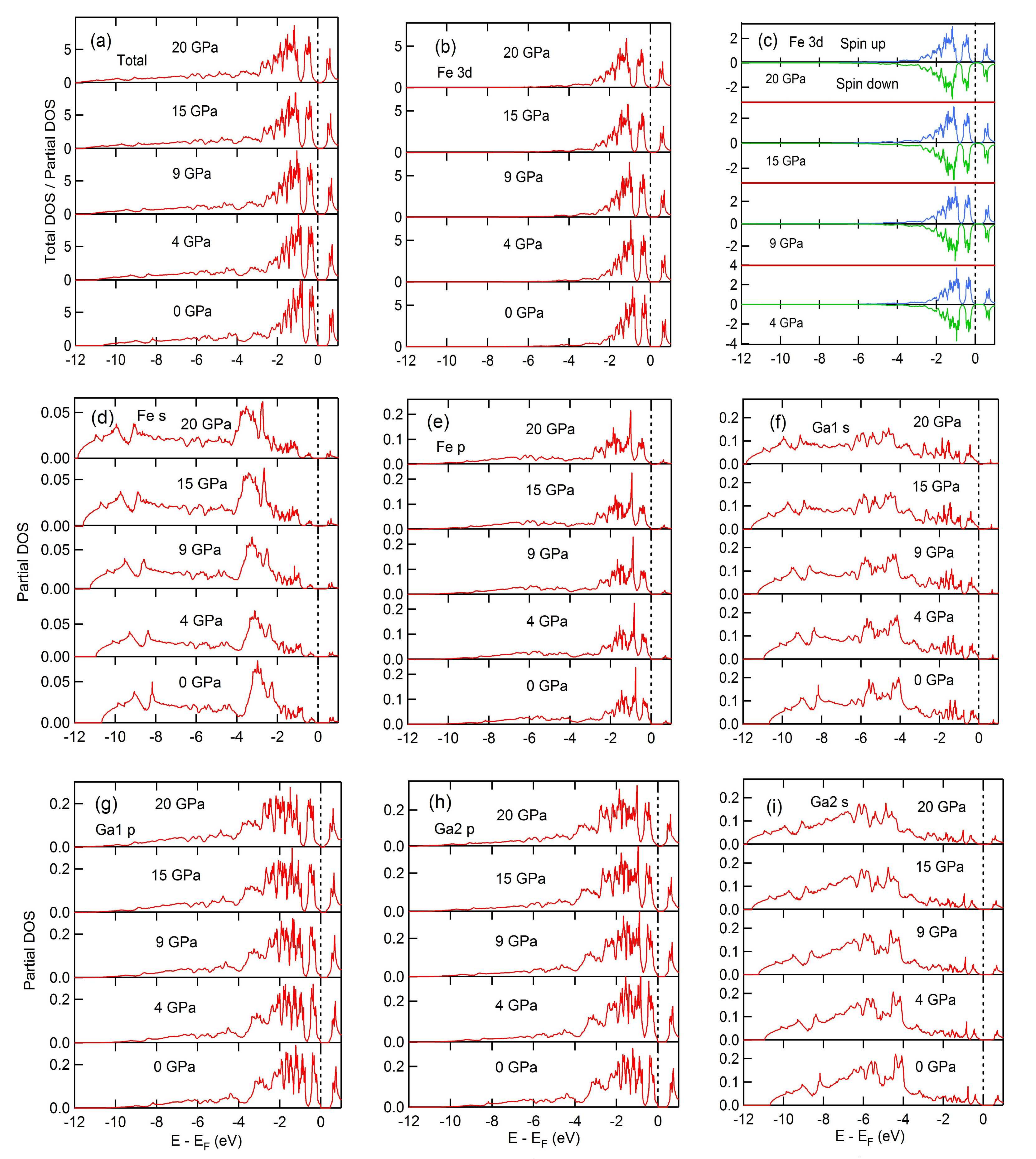}
\caption{variation of total DOS and different partial DOS of FeGa$_3$ with applied pressure for U$_{Fe(3d)}$ = 0 eV.}
\end{figure}

There are a few reports, where external pressure has been found to increase the correlation in a system. A report on CaMnO$_3$ \cite{Paszkowicz} by Paszkowicz \textit{et al.} shows that reduction in atomic size due to pressure, affects the localization of electron orbitals in Mn atom. Another report by Moreno \textit{et al.} on MnAs nanocrystal shows that the induced strain in the systems confines the d electrons and increases the on-site Coulomb repulsion enhancing the localization of 3d orbitals. \cite{Moreno} Similar type of localization has been observed in Mossbauer experiment by Kantor \textit{et al.} where it is reported that there is an increase in the hyperfine field in FeO with applied pressure and this hyperfine field is related to an increase in the magnetic moment. \cite{Kantor} In Mg$_{1-x}$Fe$_x$O \cite{Tsuchiya}, an increase in U$_{Fe(3d)}$ is observed with increase in pressure. However, the band width also increases by a greater extent, thereby pointing towards increased itinerancy of the Fe 3d electrons. The increase of band overlap and hence itinerancy with pressure has also been reported in several systems, for example, LaMnO$_3$ \cite{Zhou}, Curium metal \cite{Benedict} etc.

\begin{figure}
\centering
\includegraphics[width=0.9 \textwidth]{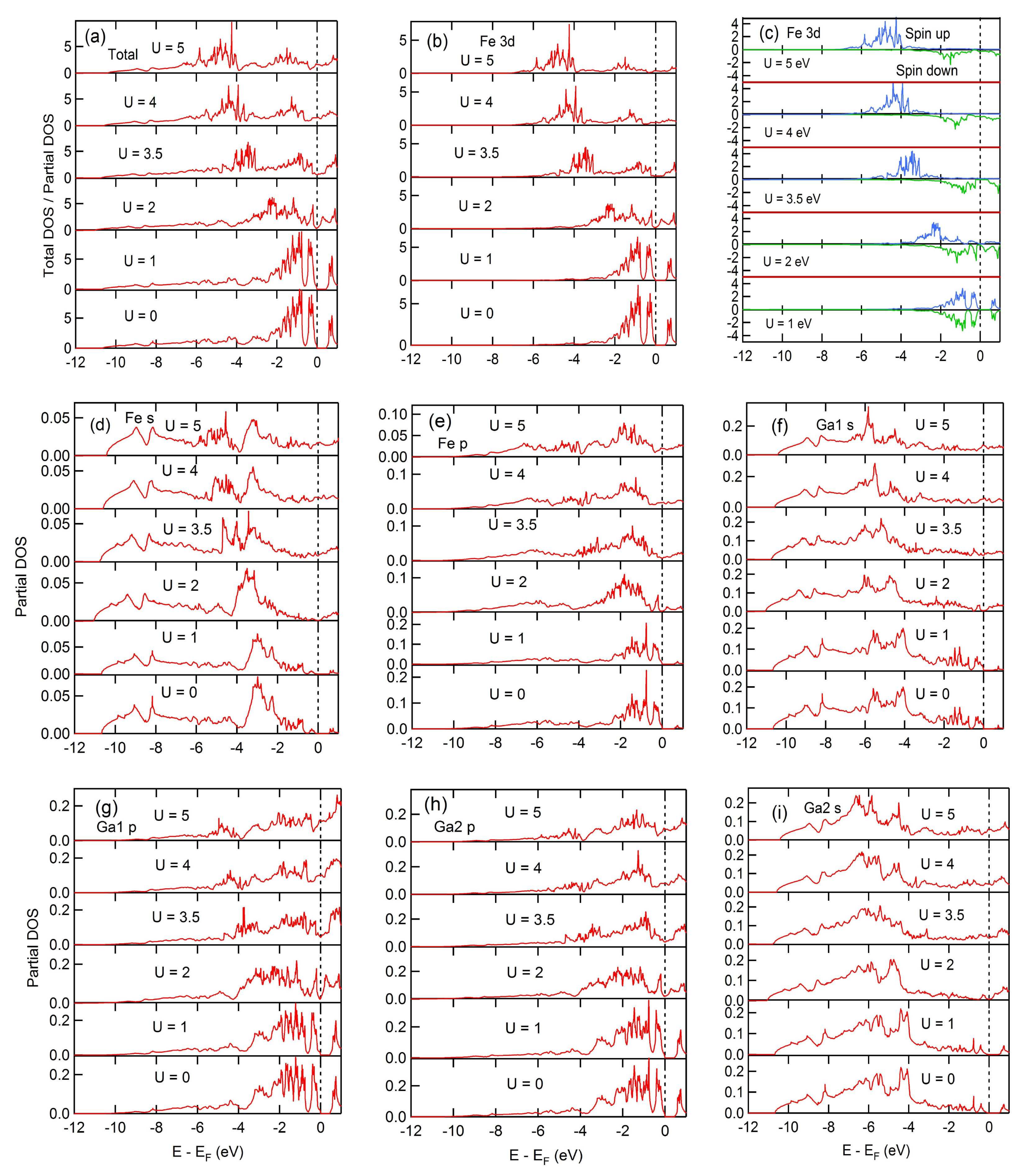}
\caption{variation of total DOS and different partial DOS of FeGa$_3$ at ambient pressure with different U$_{Fe(3d)}$. All `U' in the figure represents `U$_{Fe(3d)}$'. }
\end{figure}

The calculated electronic structures of FeGa$_3$ at different pressure with suitable U$_{Fe(3d)}$, [plotted in Fig. 5 (b)] show that the density of states get modified drastically at high pressure with larger U and the bandgap gradually decreases. We find that around a pressure of 4 GPa, a small non zero value of density of states is present at the Fermi level, which is primarily contributed from the Ga atoms. These states at the Fermi level arise from Ga atoms,  because in FeGa$_3$ the hybridization between Fe 3d and Ga 4s-4p bands is responsible for the band gap. Due to hybridization, the bands split into two parts with the lower energy band being completely filled and the upper band being empty thereby making FeGa$_3$ a semiconductor.  With the introduction of U$_{Fe(3d)}$, the Fe 3d spin up band moves away from the Fermi level and the PDOS of Fe 3d spin down states decreases near the Fermi level,  which results in a decrease of the total Fe 3d PDOS near Fermi level. This reduces the hybridization between the Fe 3d and Ga 4s-4p levels thereby forming relatively pure Ga 4s-4p states at the Fermi level. We thus conclude that in order to reproduce the experimental unit cell volume at high pressures, a finite value of U on Fe 3d electrons has to be considered. This modifies the electronic structure at high pressure [Fig. 5 (b)] by generating a small non zero density of states at the Fermi level, which is likely to lead to a decrease in electrical resistance of the material. Our observation is thus in sharp contrast to the reported semiconductor to metal transition at a much higher pressure (25 GPa) reported by Osorio-Guillen \textit{et al.}. \cite{Osorio}

\begin{figure}
\centering
\includegraphics[width=0.48 \textwidth]{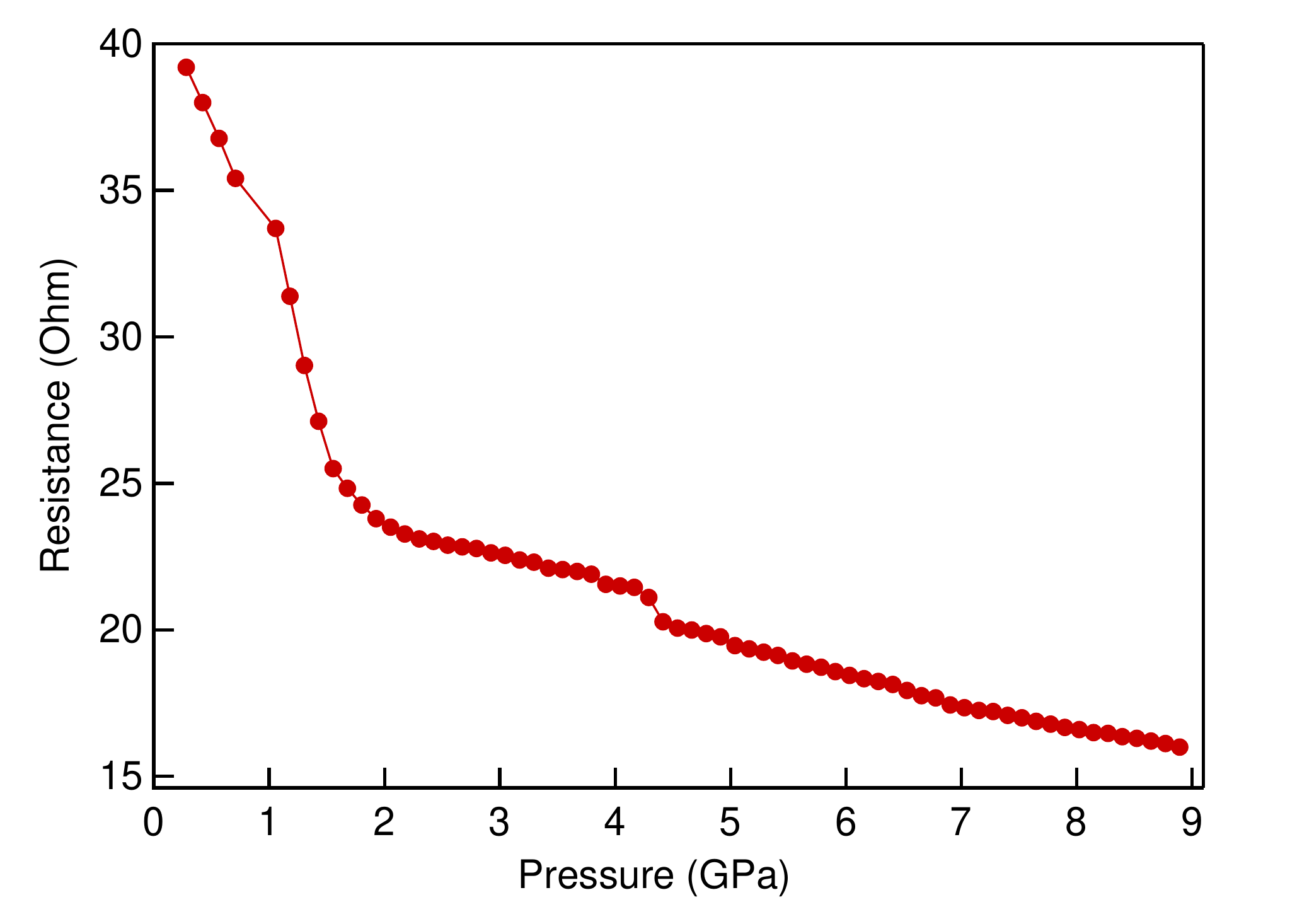}
\caption{Pressure variation of electrical resistance of FeGa$_3$.}
\end{figure}

We have further performed pressure dependent resistance measurements on FeGa$_3$ single crystal upto a maximum pressure of 9 GPa. The variation of resistance with pressure is shown in Fig. 8. In the pressure variation of resistance data the initial fall in the resistance from ambient pressure to $\sim$ 1 GPa could be attributed to a better contact formation between measuring leads and the sample since in the present experimental configuration, the contacts are formed through pressure. However beyond 1 GPa, the variation in the resistance is actually related to the behavior of the sample. We find a continuous decrease in the resistance from 1 to 9 GPa with a sharp fall in 1-2 GPa range. Beyond 2 GPa, the change in resistance with pressure is much smaller with a small discontinuity at around 4 GPa. This sharp fall in resistance around a pressure of 2 GPa is likely to be a signature of an electronic  transition in FeGa$_3$, which is observed in our calculation as well. Beyond 2 GPa, the nearly flat variation of resistance with pressure indicates that the  system is in a metallic like state. Similar kind of pressure induced electronic transition has already been observed in silicon and germanium. \cite{Alka} However, in our case the resistance of the system is higher than what we expect from a good metal. Such high values of resistance can be attributed to the following reasons: low density of states at the Fermi level, low carrier mobility due the presence of relatively flat bands, \cite{Yin} defects and impurities in the system etc.  Similar kind of high resistance has been observed in Bi. \cite {Rimas}  Thus the pressure dependent resistance measurement supports the importance of coulomb repulsion U$_{Fe(3d)}$ on the Fe 3d electron in determining the structural and electronic properties of FeGa$_3$.

\subsection{Beyond ~20 GPa pressure }

We next look at the XRD data recorded at pressures beyond 20 GPa (Fig. 2) where the intensities of the diffraction peaks corresponding to FeGa$_3$ decrease drastically and eventually vanish beyond 26.6 GPa. However, the peak intensity corresponding to Au remain almost unchanged and peak positions shift as expected with pressure. This confirms that FeGa$_3$ becomes increasingly disordered beyond  $\sim$ 20 GPa and around an applied pressure of  $\sim$ 26 GPa, FeGa$_3$ becomes completely disordered and amorphizes. The amorphous nature of FeGa$_3$ is observed upto the maximum pressure of 33.3 GPa applied in this experiment. While releasing the pressure, we find that the diffraction peaks corresponding to FeGa$_3$ are observed at the same position as in the ambient pressure data before the application of pressure. However, the intensity of the FeGa$_3$ diffraction peaks are reduced significantly and also accompanied by significant broadening.

We attribute the large disorder followed by amorphization of FeGa$_3$ to the structural instability introduced at high pressures. In the tetragonal symmetry of FeGa$_3$ the number of atoms, their arrangements, bondings and hybridization are quite different along the {\it a} and {\it c} axes, resulting in significantly different physical properties along the different lattice vectors.  As a result, hydrostatic pressure affects the different directions differently. The value of $\frac{\Delta a}{\Delta P}$ and $\frac{\Delta c}{\Delta P}$ evaluated in the pressure range of 0-20 GPa is 0.0088 \AA/GPa and 0.0127 \AA/GPa respectively. This clearly indicates that the compressibility along the xy plane is much less as compared to the z direction.

\begin{figure}
\centering
\includegraphics[width=0.45 \textwidth]{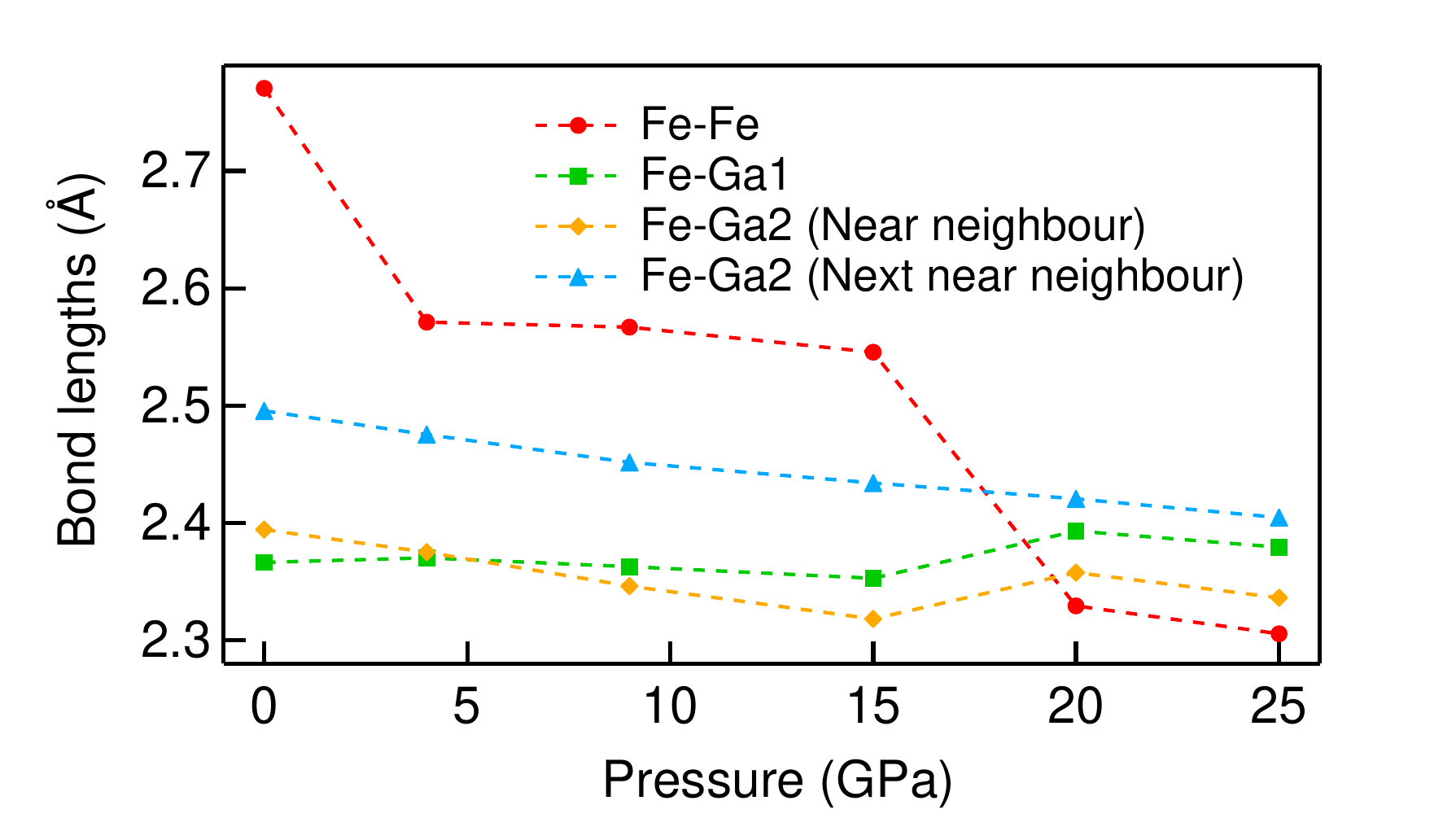}
\caption{Calculated bond lengths at different applied pressure. Dotted lines are guide to eye. Bond lengths are calculated with those U$_{Fe(3d)}$ values for which the calculated and experimental unit cell volume are the closest.}
\end{figure}

To understand the possible reason behind the observed amorphization of FeGa$_3$, we analyze the crystal structure of FeGa$_3$, based on the space group and Wyckoff position of the Fe and Ga atoms. Figure 1(b) shows the arrangement of Fe, Ga1 and Ga2 atoms in FeGa$_3$. In FeGa$_3$, the Fe atoms are aligned along the $\langle$1,1,0$\rangle$ direction forming pairs with the Fe atoms placed 2.751 \AA{} apart at ambient pressure, whereas along the $\langle$0,0,1$\rangle$ direction, these Fe pairs lie at a distance of 6.546 \AA{} from each other. In addition, Fe and Ga1 atoms lie in the same xy plane ([001] plane), at a distance of 2.370 \AA{}, whereas  there are no Fe and Ga2 atoms in the same xz or yz planes. Along lattice vector {\it c}, there are two Ga2 atoms, which lie at a distance of 2.399 \AA{} (near neighbour) and 2.498 \AA{} (next near neighbour) from the Fe atoms with different (x, y) coordinates as compared to Fe atom. This atomic arrangement makes the xy plane more stiff as compared to the xz or yz planes. On the application of pressure, the Fe and Ga atoms are expected to be displaced inside the unit cell for minimizing the energy of the system, within the restrictions posed by the $P4_2/mnm$ space group symmetry of FeGa$_3$. Accordingly, with increase of pressure, the Fe and Ga1 atoms are constrained to move along the xy plane whereas the Ga2 atoms can be displaced along the xy plane and the z directions. Figure 9 shows the variation of the calculated bond lengths: Fe-Ga1, Fe-Fe and Fe-Ga2 (near neighbor) and Fe-Ga2 (next near neighbor) with pressure. These bonds are indicated in Fig. 1(b). With application of pressure, the Fe-Ga1 and Fe-Ga2 (near neighbour) bond lengths decrease continuously with a slight increase above a pressure of 15 GPa. The Fe-Fe bond decreases throughout the pressure range (upto 25 GPa) with two sudden decrease at around 4 GPa and 20 GPa. The total change in the Fe-Fe bond length compared to ambient pressure is $\sim$ 14\%, which is an extremely large change. The sudden decrease in this bond length in the low pressure range of 0 to 4 GPa, can be possibly related to the drop of resistance in the 1-2 GPa range in the resistance versus pressure graph. Around 20 GPa, the large decrease in this Fe-Fe bond length brings the Fe-Fe atoms very close ($\sim$ 2.35\AA), which is less than the diameter of atomic Fe at ambient pressure. Further increase in the applied pressure tend to move the Fe atoms out of the [001] plane, thereby breaking the symmetry of the system. This makes the system disordered and it eventually amorphizes. The small increase in the Fe-Ga1 and Fe-Ga2 (near neighbour) bond distance above 15 GPa pressure is actually an effect of large decrease in Fe-Fe bond length. The other Fe-Ga2 (next near neighbor) bond length decreases monotonically with increasing pressure.

\section{Conclusion}
High pressure X-ray diffraction measurements (XRD) on FeGa$_3$ have been performed from ambient pressure to $\sim$ 33 GPa. It has been observed that the system remains in its tetragonal structure below a pressure of  $\sim$ 20 GPa. In the pressure range of 0 to 20 GPa we have found a mismatch between the experimental and calculated equation of state at high pressure and this mismatch increases with increasing pressure. To check the origin of the mismatch, we have also performed calculation with GGA+U method. Results of our calculations indicate that, certain value of U$_{Fe(3d)}$ at particular pressure reproduces the experimental EOS and the value of required U$_{Fe(3d)}$ increases as pressure increases. However, the increase of U$_{Fe(3d)}$ at high pressure does not indicate that the system is strongly correlated, as the band width also increases in the same proportion. Electronic structure calculations show that around a pressure of 4 GPa a small finite density of states arise at the Fermi level, and a signature of this phenomenon has been observed in our pressure dependent resistance measurements as well. Above $\sim$ 26 GPa pressure, FeGa$_3$ becomes completely disordered and amorphoizes. Upon the release of pressure, the diffraction peaks are re-observed but with much lower intensity and larger width.

\section{Acknowledgement}

The authors thank Shri Ajay Kak for the active support in preparing the sample for the experiment, and Shri V. K. Ahire for the experimental support. Dr. P. A. Naik is cordially thanked for valuable scientific discussion and encouragement. CK and AC thank Scientific Computing Group,Computer Division, RRCAT for their support. DM also would like to thank RRCAT for financial support.


\begin{thebibliography}{}
\bibitem{Jaccarino} V. Jaccarino, J. H. Wernick, L. R. Walker and S. Arajas, Phys. Rev. {\bf 60}, 476 (1967).
\bibitem{Petrovic} C. Petrovic, J. W. Kim. S. L. Budko, A. I. Goldman, P. C. Canfield, W. Choe, and G. J. miller Phys. Rev. B {\bf 67}, 155205 (2003).
\bibitem{Weinert} M. Weinert and R. E. Watson. Phys. Rev. B {\bf58}, 15 (1998);
\bibitem{Amagai} Y. Amagai, A. Yamamoto, T. Iida and Y. Takanashi,  Journal of Appl. phys. {\bf 96}, 5644 (2004).
\bibitem{Bogdanov} D. Bogdanov, K. Winzer, I. A. Nekrasov and T. Pruschke, J. Phys. Condens. Matter {\bf 19}, 232202 (2007).
\bibitem{Bentien} A. Bentien, S. Jonsen, G. K. H. Madsen, B. B. Iversen and F. Steglich , Europhrs. Lett. {\bf 80}, 17008 (2007).
\bibitem{Goncalves} L. M. Goncalves, C. Couto, P. Alpuim, A. G. Rolo, F. Volklein, J. H. Correia, Thin Solid Films {\bf 518}, 2816 (2010).
\bibitem{Hadano} Y. Hadano, S. Narazu, M. A. Avila, T. Onimaru, and T. Takabatake, J Phys. Soc. Japan, {\bf 78}, 013702 (2009).
\bibitem{Kasinathan} D. Kasinathan,K. Koepernik and H. Rosner, Phys. Rev. B {\bf 85}, 035207 (2012);
\bibitem{Haldolaarachchige} N. Haldolaarachchige, A. B. Karki, W. Adam Phelan, Y.M. Xiong, R. Jin, Julia Y. Chan, S. Stadler and D.P. Young,  J. Appl. Phys. {\bf 109}, 103712 (2011).
\bibitem {Takagiwa} Y. Takagiwa, K. Kitahara, Y. Matsubayashi, and K. Kimura, J Appl. Phys. {\bf 111}, 123707 (2012).
\bibitem{Manyala} N. Manyala J. F. DiTusa G. Aeppli and A. P. Ramirez, Nature, {\bf 454}, 976 (2008).
\bibitem{Umeo} K. Umeo, Y. Hadano, S. Narazu, T. Onimaru, M. A.  Avila, and  T. Takabatake, Phys. Rev. B {\bf 86}, 144421 (2012).
\bibitem{Fu} C. Fu, M. P. C. M. Krijn, S. Doniach, Phys. Rev. B, {\bf 49}, 2219 (1994).
\bibitem{Arita2} M. Arita, K. Shimada, Y. Takeda, M. Nakatake, H. Namatame, M. Taniguchi, H. Negishi, T. Oguchi, T. Saitoh, A. Fujimori, and T. Kanomata, Phys. Rev. B {\bf 77}, 205117 (2008).
\bibitem{Herzog} A. Herzog, M. Marutzky, J. Sichelschmidt, F. Steglich, S. Kimura, S. Johnsen and B. B. Iversen, Phys. Rev. B {\bf 82}, 245205 (2010).
\bibitem{Tsujii} N. Tsujii, H. Yamoaka, M. Matsunami, R. Eguchi, Y. Ishida, Y. Senba, H. Ohashi, S. Shin, T. Furubayashi, H. Abe and H. Kitazawa, J. Phys. Soc. Jpn. {\bf 77}, 024705 (2008).
\bibitem{Arita} M. Arita, K. Shimada, Y. Utsumi, O. Morimoto, H. Sato, H. Namatame, M. Taniguchi, Y. Hadano and T. Takabatake, Phys. Rev. B {\bf83}, 245116 (2011).
\bibitem{Ulrich2} U. Haussermann, M. Bostrom, P. Viklund, O. Rapp and T. Bjornangen, Journal of Solid State Chemistry {\bf 165}, 94 (2002).
\bibitem{Imai} Y. Imai, A. Watanabe, Intermetallics {\bf 14}, 722 (2006).
\bibitem{Yin} Z. P. Yin and W. E. Pickett, Phys. Rev. B {\bf 82}, 155202 (2010).
\bibitem{Osorio} J. M. Osorio-Guillén, Y. D. Larrauri-Pizarro and G. M. Dalpian, Phys. Rev. B {\bf 86}, 235202 (2012).
\bibitem{Gamza} M. B. Gamza , J. M. Tomczak, C. Brown, A. Puri, G. Kotliar and M. C. Aronson, Phys. Rev. B {\bf 89}, 195102 (2014).
\bibitem{Storchak} V. G. Storchak, J. H. Brewer, R. L. Lichti, R. Hu, and C. Petrovic, J. Phys.: Condens. Matter. {\bf 24}, 185601 (2012).
\bibitem{phaseDia} Fe-Ga Binary Alloy Phase Diagrams, II Ed., Ed. T.B. Massalski, {\bf 2}, 1702-1704 (1990).
\bibitem{Pandey} K. K. Pandey , H. K. Poswal, A. K. Mishra, A. Dwivedi, R Vasanthi, N.  Garg, S. M. Sharma, Pramana {\bf 80}, 607 (2013).
\bibitem{Kresse1} G. Kresse and J. Furthmuller, Phys. Rev. B {\bf 54}, 11169 (1996).
\bibitem{Kresse2} G. Kresse and D. Joubert, Phys. Rev. B {\bf 59}, 1758 (1999); VASP 5.2 programme package is fully integrated in the Mede A platform (Materials Design, Inc.) with a graphical user interface enabling the computation of the properties.
\bibitem{Perdew} J. P. Perdew, K. Burke and M. Ernzerhof, Phys. Rev. Lett. {\bf 77} 3865 (1996).
\bibitem {Liechtenstein} A. I. Liechtenstein, V. I. Anisimov, and J. Zaanen,  Phys. Rev. B 52, {\bf R5467} (1995).
\bibitem{Loschen} C. Loschen, J. Carrasco, Konstantin M. Neyman and F. Illas, Phys. Rev. B {\bf 75}, 035115 (2007).
\bibitem{Trimarchi}G. Trimarchi and N. Binggeli, Phys. Rev. B {\bf 71}, 035101 (2005).
\bibitem{Fang} Z. Fang, I. V. Solovyev, H. Sawada and K. Terakura, Phys. Rev. B {\bf 59}, 762 (1999).
\bibitem{Weibing} Wei-Bing Zhang, Yu-Lin Hu, Ke-Li Han and Bi-Yu Tang,  J Phys Cond Matter. {\bf 18}, 9691 (2006).
\bibitem{Asok} A. K. Verma, P. Modak, S. M. Sharma, A. Svane, N. E. Christensen, and S. K. Sikka, Phys. Rev. B {\bf 88}, 014111 (2013).
\bibitem {Arroyo} M. E. Arroyo-de Dompablo, A. Morales-Garcia, and M. Taravillo,  J. Chem. Phys. {\bf 135}, 054503 (2011).
\bibitem{Anita} A. S. Botana, Y. Quan and W. E. Pickett, {\bf arxiv 1509.08517v2} [cond-mat.str-el] (30 Nov 2015).
\bibitem{Tsuchiya} T. Tsuchiya, R. M. Wentzcovitch, C. R. S. Silva, and S. Gironcoli, Phys. Rev. Lett. {\bf 96}, 198501 (2006).
\bibitem{Paszkowicz} W. Paszkowicz, S. M. Woodley, P. Piszora, B. Bojanowski, J. Pietosa, Y. Cerenius, S. Carlson and C. Martin, Appl. Phys. A {\bf 112}, 839 (2013).
\bibitem{Moreno} M. Moreno, J. I. Cerda, K. H. Ploog, and K. Horn, Phys. Rev. B {\bf 82}, 045117 (2010).
\bibitem{Kantor} A. P. Kantor, S. D. Jacobsen, I. Yu. Kantor, L. S. Dubrovinsky, C. A. McCammon, H. J. Reichmann, and I. N. Goncharenko, Phys. Rev. Lett. {\bf 93}, 215502 (2004).
\bibitem{Zhou}J .S. Zhou and J. B. Goodenough, Phys. Rev. Lett. {\bf 89}, 087201 (2002).
\bibitem{Benedict} U. Benedict, R. G. Haire, J. R. Peterson and J. P. Itie, Journal of Physics F: Metal Physics {\bf 15}, L-29 (2000).
\bibitem{Alka} A. B. Garg, V. Vijayakumar and B. K. Godwal, Review of Scientific Instruments {\bf 75}, 2475 (2004).
\bibitem{Rimas} J. R. Vaisnys and R. S. Kirk, Journal of Applied Physics {\bf 38}, 4335 (1961).
\end{thebibliography}
\end{document}